# Role of self-coherence in single-electron phase contrast imaging

C. Kisielowski[1,2,*], P. Specht[2,3], J.R. Jinschek[4], S. Helveg[1,†]

1 Center for Visualizing Catalytic Processes (VISION), DTU Physics, Technical University of Denmark, DK-2800 Kgs. Lyngby, Denmark
2 Electron Scattering Solutions, Piedmont CA, 94611, USA
3 Department of Materials Science and Engineering, University of California Berkeley, Berkeley, CA 94720, USA
4 Center for Visualizing Catalytic Processes (VISION), DTU Nanolab. Technical University of Denmark, DK-2800 Kgs. Lyngby, Denmark

**ABSTRACT**. The extension of coherent lattice contrast into the energy loss region in high-resolution transmission electron microscopy (HRTEM) is described by a pulse-like electron-sample interaction in the energy/time uncertainty limit $\Delta E\Delta t \simeq \hbar/2$. It generates a wave packet by electron self-interference in any coherent-inelastic scattering event with energy loss $\Delta E$. The width of this wave packet is characterized by a self-coherence length $l_s(\Delta E)$ that is predictable because an intrinsic decoherence phase $\Delta\varphi$ around one radian is set by the expectation value $\langle\exp(i\Delta\varphi)\rangle$ for phase fluctuations. In this case the visibility of interference contrast from a crystalline sample with lattice parameter $a$ is limited by a Rayleigh-like transfer factor $P(l_s, a)$ in the self-coherently illuminated sample area. The model is verified by energy-filtered HRTEM images of hexagonal BN and identifies energy-loss-induced phase noise as a single-electron visibility limit distinct from resolution limitations caused by ensemble-coherence or counting-statistical noise.

A century ago, E. Schrödinger formulated his differential equation describing the propagation of a single electron in matter by a quantum-mechanical wave function, establishing a cornerstone of modern quantum physics and electron microscopy [1,2]. A particularly fundamental aspect of quantum physics is self-coherence, which refers to the ability of a single electron wave function to interfere with itself. It results in the formation of a wave packet from an infinite wave function, as already pointed out by Schrödinger himself in his attempt to describe the transition from micro- to macro-mechanics [3]. In fact, self-coherence is implicitly incorporated in solutions of Schrödinger's static equation, using e.g. multislice simulations [4,5]. Beyond this computational role, however, self-coherence has rarely been considered as a distinct physical quantity contributing to image formation in high-resolution transmission electron microscopy (HRTEM).

Instead, coherence in HRTEM is conventionally described in terms of interference between ensembles of transmitted and scattered electrons [6,7]. Within this framework, image contrast is typically associated with elastic or quasi-elastic scattering processes, while spatial and temporal coherence of the electron source [8], residual lens aberrations [9] and counting statistics [10], determine the attainable image contrast from state-of-the-art equipment. Yet, energy-filtered HRTEM experiments have demonstrated that coherent lattice contrast persists deep into the electron energy-loss regime, vanishing only at energy losses of several hundred electron volts even in thin samples [11-13]. The origin of this remarkable persistence of coherent phase contrast following inelastic electron-matter interactions remains unresolved.

Seeking resolve, energy-filtered HRTEM measurements have recently revealed a close correspondence between the electron self-coherence length or self-coherence time and the persistence of atomic-resolution contrast as a function of energy loss, $\Delta E$, in energy-filtered HRTEM (Fig. 1) [13-15]. Relating to Schrödinger's time-dependent equation, an electron-sample interaction was assumed to be an inelastic pulse-like event in the energy/time uncertainty limit $\Delta E\Delta t \simeq \hbar/2$, where $\hbar$ is the reduced Planck constant and $\Delta t$ the related time uncertainty that can be interpreted as an interaction time. Moreover, Fig. 1 shows that a time scale of femtoseconds to attoseconds is mapped onto a nanometer length scale in this process because electrons travel close to the speed of light c in HRTEM experiments. This length scale is readily accessibly by HRTEM because of its high spatial resolution (~ 0.5 Å) and small electron wavelength (~ 2 pm). Since the measurement detected one electron at a time [14], the finding suggest that the coherent lattice transfer is governed not simply by the magnitude of the energy loss but by the ability of the scattered electron wave function to maintain the capability for self-interference. The fundamental mechanism limiting electron self-coherence, however, remains unknown.

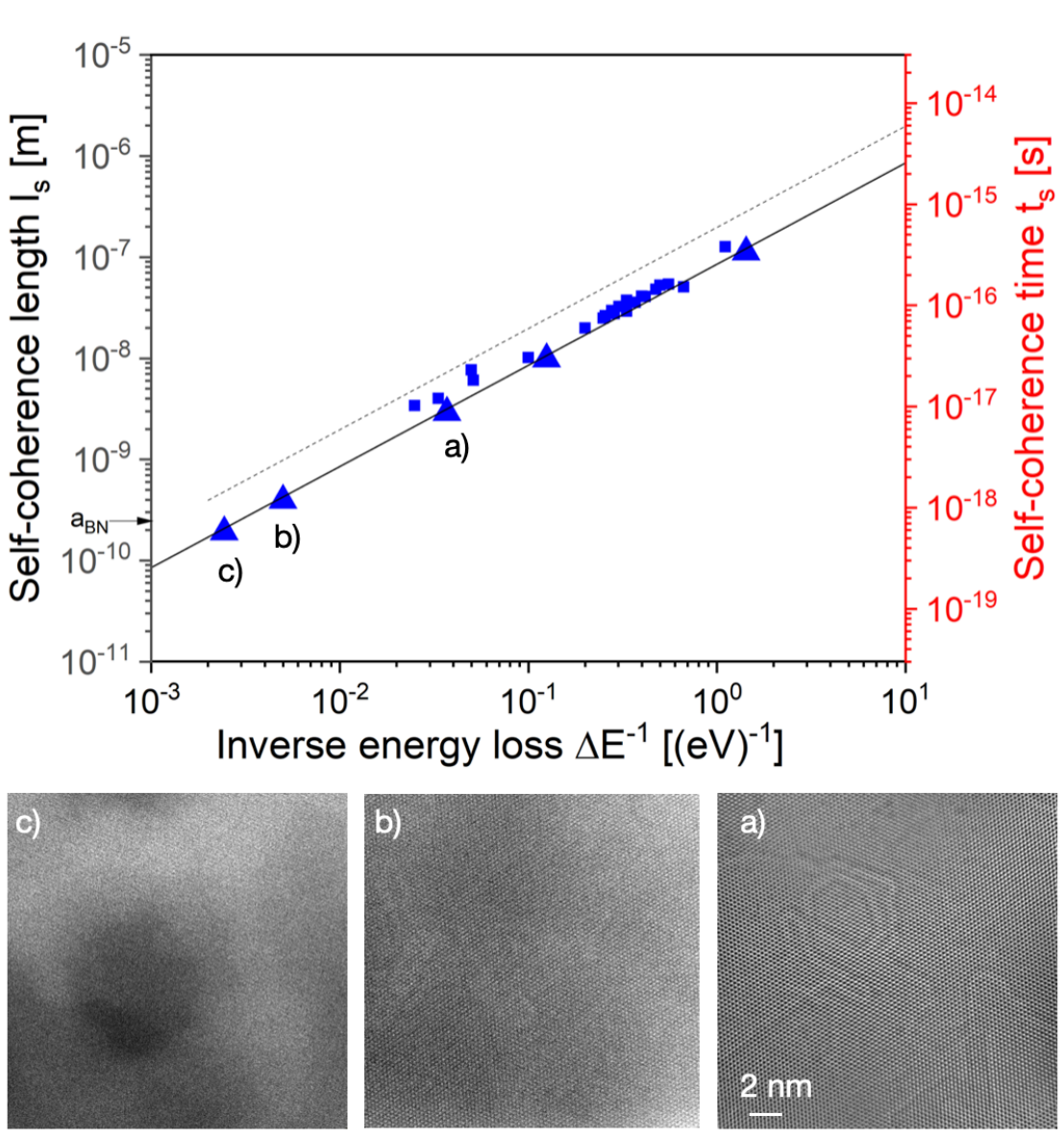

FIG. 1. Self-coherence time $t_s$ (right ordinate) and the related self-coherence length $l_s = t_s\, c$ (left ordinate) as functions of energy loss $\Delta E$. The solid line is the energy/time uncertainty limit calculated from wave length differences for a decoherence phase of 0.5 rad [14,7]. The dotted curve is obtained by Eq. (2) for a decoherence phase of one radian (= 1 rad) [13,7]. Blue squares are $l_s$ measurements [15]. The blue triangles mark energy losses and the related $l_s$ values where energy-filtered HRTEM images were acquired and a) - c) are selected examples [13, 16]. Note that the interference lattice contrast vanishes as $l_s$ becomes comparable to the crystal lattice periodicity $a$.

Here we examine electron self-coherence measurements by a quantitative model that relates inelastic energy loss $\Delta E$, self-coherence length $l_s$, and contrast transfer through a cumulative probability distribution sensitive to both amplitude and phase fluctuations. The model identifies an intrinsic decoherence phase of one radian (1 rad) as a fundamental limit for self-coherence and accounts for the persistence of linear interference contrast in the inelastic scattering regime up to energy losses of several hundred electron volts. These results establish electron self-coherence as an experimentally accessible observable and reveal quantum limits for contrast visibility in electron phase-contrast imaging.

*Contact author: cfkisielowski@sbcglobal.net, †Contact author: stig@fysik.dk

We note that in the formation process of a wave packet by self-interference [7] phase fluctuations $\Delta\varphi$ will unavoidably occur with an expectation value $\langle\exp(i\Delta\varphi)\rangle$ in the limit of the energy / time uncertainty. If $\Delta\varphi$ is described by a zero-mean Gaussian distribution $P(\Delta\varphi)$ with variance $\sigma_\varphi^2 = \langle\Delta\varphi^2\rangle$, then

$$\langle\exp(i\Delta\varphi)\rangle = \int \exp(i\Delta\varphi)\, P(\Delta\varphi)\, d\Delta\varphi = \exp(-\sigma_\varphi^2/2). \quad (1)$$

Thus, for $\sigma_\varphi^2 = 1\ \text{rad}^2$ the amplitude of a wave packet reduces to $1/\sqrt{e} \approx 0.61$. This value acts as an operational marker for substantial decoherence, not as a sharp coherence cutoff and explains why a critical value of 1 rad appears in previous derivations [e.g. 17, 18] and measurements of the self-coherence length [13]. Herein the self-coherence length was related to an electron energy loss as [7]:

$$l_s = \lambda^2/\Delta\lambda \approx \hbar c/\Delta E \quad (2)$$

$l_s(\Delta E)$ is interpreted as the width of a wave packet and can be used to estimate a self-coherence volume $l_s(\Delta E)^3$ wherein self-interference remains possible for any energy loss $\Delta E$.

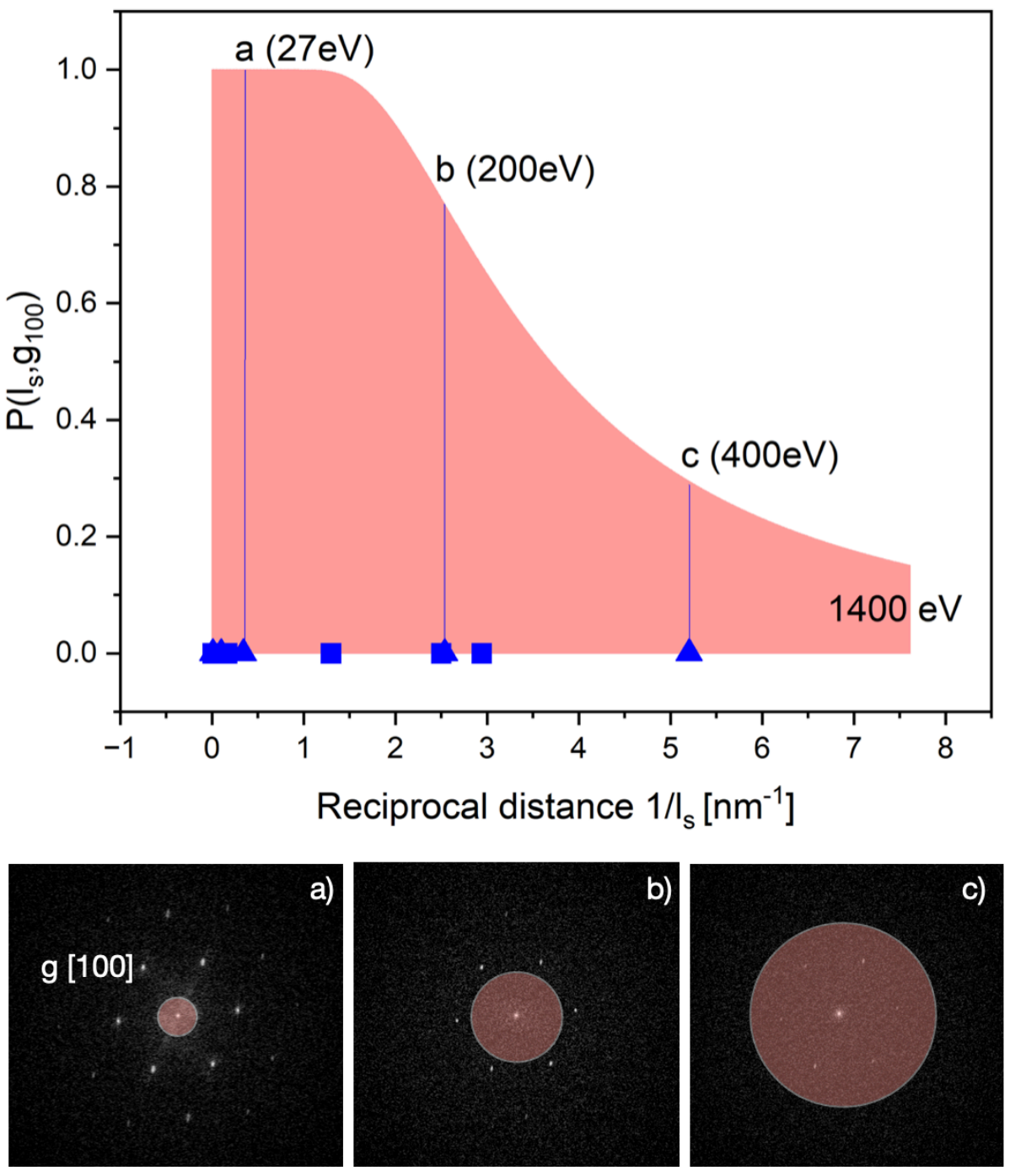

FIG. 2. Calculated cumulative Rayleigh-type coherence transfer probability $P(l_s, g_{100})$ of BN for energy losses $\Delta E$ up to 1400 eV ($l_s$ = 0.13 nm). Selected energy losses are highlighted that refer the energy filtered HRTEM images from Fig. 1. Their Fourier transform is shown as insets where the presence of a self-coherence length $l_s$ is indicated as a opaque virtual aperture of radius $R = l_s$. For details see text.

Concerning the wave amplitude $A$, its contribution to the visibility of interference contrast depends on the number $N$ of coherently illuminated periodic units as known from e.g. diffraction gratings, $A \sim N$. For a two dimensional description of interference transfer from a self-coherently illuminated area, the Rayleigh distribution is employed. In this case, two independent but identical normal distributions of mean zero are assumed in two dimensions to give the cumulated probability distribution:

$$P(r,s) = 1 - \exp[-(r/s)^2 / 2]. \quad (3)$$

Where $A = |r| = (A_x^2+A_y^2)^{1/2}$ is the modulus of the amplitude and $s^2$ is their common scale. Setting $r = l_s$ and $s = a$ (lattice parameter) it is clear that their ratio $(l_s/a)^2$ estimates the number $N$ of lattice unit cells that are self-coherently illuminated. Self-interferences must vanish if constructive interference is no longer possible because $N \leq 1$. However, this boundary is "soft" because decoherence adds uncertainty to $l_s$ (Fig. 1). In the other limit, a largest $l_s$ exists because there is energy exchange in any coherent-inelastic scattering process, which keeps the self-coherently illuminated area finite. Therefore, equation (3) is a statistical visibility criterion for the observation of self-interferences.

Fig. 2 shows that the abrupt loss of interference contrast is correctly predicted to occur for the lattice images recorded at energy losses $\Delta E$ of 200 eV and 410 eV (Fig.1) [16] by the self-coherence model of Eq. (3). Reciprocal distances $1/l_s$, are employed to highlight this energy range and the lattice parameter $a$ = 0.26 nm of BN is represented by its detected lattice spatial frequency $g_{100}$ = 4.4 nm$^{-1}$ which corresponds to a real-space periodicity of approximately 0.23 nm. In this presentation $P(l_s,g_{100}) = 1$ is the cumulated probability that constructive self-interferences can occur because the self-coherently illuminated area remains large enough. Fig. 2 is generated by only using the lattice parameter of BN and computed self-coherence lengths from either Eq. (2) or the time/energy uncertainty as shown by the straight lines in Fig 1. They represent decoherence phases of 0.5 and 1 rad. $P(l_s, g_{100})$ was calculated from equation (3). Note that the experimentally observed data are also indicated in Fig. 2. The model does not suggest that electrons are no longer detected at high energy loss. Instead, it predicts that the detected electrons increasingly cease to contribute to constructive self-interferences associated with a specific spatial frequency. Linear lattice contrast persists when $P(l_s, g_{100})$ approaches unity and vanishes when $l_s$ becomes comparable to the lattice periodicity. If suppressed, residual coherent contrast may still encompass nonlinear $g$–$g$ interferences [19], which is material independent and was observed in BN and STO [11, 12]. Multiple scattering events are absent in our sample due to its small thickness of only a few monolayers of BN. Consequently, the proposed energy dependent self-coherence factor introduces a single-electron transfer criterion to conventional HRTEM interpretation.

The effect of a self-coherence length on the transmission of spatial lattice frequencies can be visualized by an opaque virtual aperture of radius $1/l_s$ in the Fourier transform of lattice images as shown in Fig. 2. It is convenient to compare this altered transfer to the established Contrast Transfer Function (CTF) of advanced aberration-corrected electron microscopes [20, 21]. The Fig. 3 highlights that this comparison leads to a clear cut distinction between self-coherence and ensemble coherence. At low energy loss, $P(l_s,g)$ approximates 1 and does not affect relevant lattice frequencies. Consequently, ordinary phase-contrast interpretation remains valid even if energy losses occur during scattering (Fig. 3a). Furthermore, the established resolution limitation imposed by ensemble coherence remains unchanged. However, at energy loss above ~200 eV, $l_s$ diminishes, and $P(l_s, g_{100})$ initially suppresses the direct transfer of low spatial lattice frequencies (Fig. 3b), even though electrons continue to be detected by the detector and contribute to an incoherent background. Since a Nelsonian illumination scheme is used in the HRTEM experiments [8,14,21] the image of the electron source is directly focused onto the sample. This allows matching the irradiated area and the self-coherently illuminated area for an optimal contrast transfer at a minimum of electron irradiation damage.

For amorphous specimens, the same dependence should be interpreted as an energy-loss-dependent envelope on the diffuse spatial-frequency spectrum rather than as a cutoff for discrete Bragg reflections. In this scenario, contrast is not expected to abruptly vanish at a discrete reflection; instead, the diffuse spectrum should be gradually re-weighted. Consequently, the self-coherence function is an additional energy-loss-dependent transfer factor, distinct from conventional source-coherence and aberration envelopes.

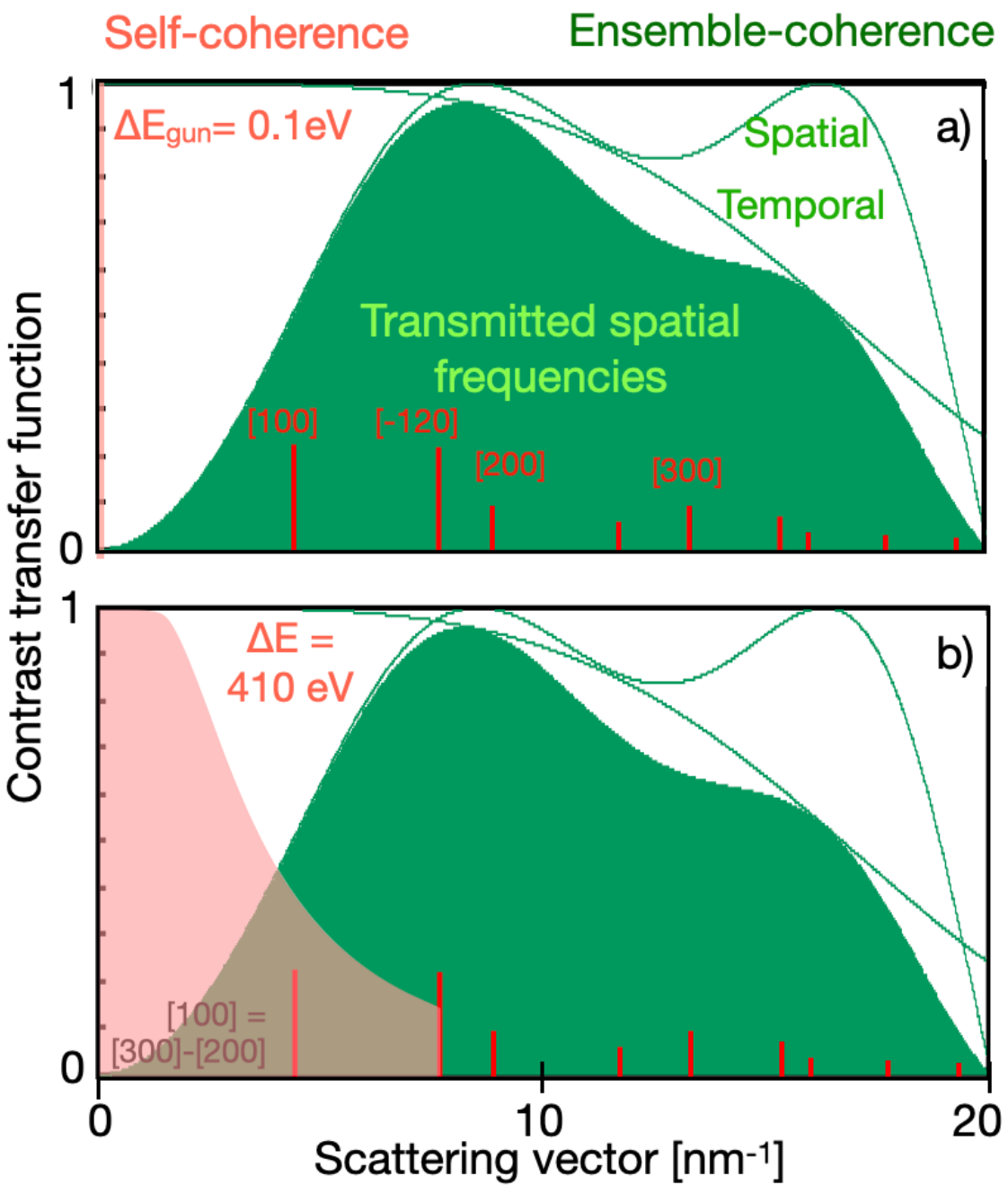

FIG. 3. Effect of self-coherence on the Contrast Transfer Function (CTF). a) At low energy losses $\Delta E$, self-coherence does not limit transfer. Spatial and temporal ensemble coherence damping envelopes are shown that set the microscope resolution. b) At high loss direct transfer of $g_{100}$ is suppressed. In this case residual contrasts can only arise from weak non-linear pathways. A non-linear *g-g* excitation ($g_{100}$ = $g_{300}$ - $g_{200}$) is indicated. Aberration parameters for 300 keV: defocus $\Delta f$ = 5nm, spherical aberration $C_3$ = -10 µm and $C_5$ = 3.5 mm with energy spread $\Delta E_{\mathrm{gun}}$ = 0.1 eV and focus spread = 0.8 nm,

Previous coherence length measurements usually refer to ensemble-coherence. H. Lichte and M. Lehmann reported on a specially designed holographic experiment to measure the coherence of electron ensembles scattered in an electric field [22] assuming a coherent-elastic interaction. They report on an ensemble-coherence length of 260 nm for 100 keV electrons. Choosing a typical energy spread of a field emission electron gun $\Delta E_{\mathrm{gun}}$ = 0.7 eV, we estimate using Eq. (2) a self-coherence length of 270 nm for $\sigma_\varphi$ = 1 rad. The close agreement of our data shows that a distinction from ensemble-coherence is challenging if energy losses are smaller than the energy spread of the electron source $\Delta E$ < $\Delta E_{\mathrm{gun}}$.

To our knowledge, Uhlemann et al. reported a smallest energy loss for electrons interacting with the conductive parts of an aligner tube in a TEM by accumulating magnetic field noise from thermally driven currents [23]. Estimating the related energy loss by $\Delta E$(1K) = 8.6 $10^{-5}$ eV the self-coherence length could reach a mm range. Related, Kerker et al. [24] reported on Coulomb interactions above sample surfaces that yield ensemble coherence values in the range of µm using a 1 keV electron beam in a specialized holographic set-up. Thus, electrons are inelastically scattered in these experimental environments suggesting that electron scattering is always coherent-inelastic. Surely, the coherent-elastic approximation remains a very useful tool to describe scattering processes with $\Delta E$ < $\Delta E_{\mathrm{gun}}$ but the approximation seems strained if energy differences of $10^{-15}$ eV are discussed. [5, 22].

A finite coherence length of inelastically scattered electrons is also known from capturing holographic diffraction patterns in the energy loss region up to $\Delta E$ = 400 eV [25]. In this case, the localization of a wave function by the formation of a wave packet results in substantial broadening of the diffraction peaks. This observation aligns with our description of self-coherence as a virtual aperture in the low spatial frequency range. However, it remains challenging to distinguish between linear and non-linear interferences in diffraction mode, particularly when samples are thick and multiple scattering events occur.

In summary, we have shown that the loss of linear lattice contrast in energy-filtered HRTEM can be described by a single-electron self-coherence model. An inelastic, pulse-like energy loss in the energy/time uncertainty limit defines a finite self-coherence length, while Gaussian phase fluctuations generate a decoherence phase of ~ one radian according to Eq. (1). Inelastic energy loss defines a finite self-coherence length through Eq. (2). By treating the resulting self-coherent wave packet as an effective transverse coherence aperture, the Rayleigh-type cumulative transfer probability of Eq. (3) is obtained. Applied to BN, this framework accounts for the disappearance of linear lattice contrast at $\Delta E$ = 200 - 400 eV. The result identifies energy-loss-induced phase noise as a single-electron contribution to contrast-transfer limits, distinct from counting-statistical noise and conventional ensemble-coherence damping. Further it strengthens the relevance of self-interferences in single electron scattering events.

The model is intentionally modest in scope to keep it transparent but specific enough to highlight the importance of time-dependent electron scattering. The pivotal model assumption is that the electron-sample interaction occurs during a pulse-like interaction in the energy/time uncertainty limit. Since the related uncertainties are set by general properties of a Fourier Transform and are fluctuations, the existence of a decoherence phase becomes a mathematical necessity that points to the origin of the energy/time uncertainty. Moreover, the approach provides an experimentally testable transfer criterion for coherent-inelastic image formation.

**ACKNOWLEDGMENTS**

The Center for Visualizing Catalytic Processes is sponsored by the Danish National Research Foundation (DNRF146). CK acknowledges fruitful discussions with Dirk van Dyck.

[1] E. Schrödinger, Phys. Rev. 28, 1049 (1926).

[2] Schrödinger's equation is non-relativistic, which is why we utilize non-relativistic electron wave lengths.

[3] E. Schrödinger, Naturwissenschaften, 664-666 (1926).

[4] J.M. Cowley, F.M. Moody, Acta Cryst 10, 609 (1957).

[5] D. Van Dyck, H. Lichte, and J.H.C. Spence, Ultramicroscopy 81, 187–194 (2000).

[6] e.g. C.B. Carter and D.B. Williams, 2016, eds, Transmission Electron Microscopy, (Springer International Publishing, Switzerland)

[7] See SM [url] for properties of self-coherence and ensemble-coherence.

[8] P.C. Tiemeijer, M. Bischoff B. Freitag, and C. Kisielowski, Ultramicroscopy 114, 72–81 (2012).

[9] M. Haider, P. Hartel, H. Müller, S. Uhlemann, J. Zach, Microsc. Microanal. 16, 393–408, (2010).

[10] C. Dwyer, Phys. Rev. Lett. 130, 056101 (2023).

[11] A. Howie, Ultramcroscopy 111 (2011) 761–767

[12] B.D. Forbes, L. Houben, J. Mayer, R.E. Dunin-Borkowski, L.J. Allen, Ultramicroscopy 147, 98–105 (2014)

[13] C. Kisielowski, P. Specht, J.R. Jinschek and S. Helveg, Microscopy and Microanalysis 31, ozae107 (2025)

[14] C. Kisielowski et al., Microscopy and Microanalysis 27, 1420–1430 (2021).

[15] C. Kisielowski, P. Specht, S. Helveg, F.-R. Chen, B. Freitag, J. Jinschek, D. Van Dyck, Nanomaterials 13, 971 (2023).

[16] See SM [url] for the complete series of lattice images recorded at the different energy losses.

[17] D. Halford, Seminar on Frequency Standards and Metrology, Quebec 1971, https://tf.nist.gov/general/pdf/33.pdf

[18] A. Jechow, B.G. Norton, S, Händel, V. Blüms, E.W. Streed, and D. Kielpinski, Phys. Rev. Lett. 110, 113605 (2013).

[19] S. Bals, R. Kilaas, and C. Kisielowski, Ultramicroscopy 104, 281–289 (2005).

[20] N. Alem et al., Phys. Rev. Lett. 106, 126102 (2011).

[21] I. Brian et al., Ultramicroscopy 282, 114328 (2026).

[22] H. Lichte, and M. Lehmann, Rep. Prog. Phys. 71, 016102 (2008).

[23] S. Uhlemann, H. Müller, P. Hartel, J. Zach, and M. Haider, Phys. Rev. Lett. 111, 046101 (2013).

[24] N. Kerker, R. Röpke, L.M. Steinert, A Pooch, A Stibor, New J. Phys. 22063039 (2020).

[25] R.A. Herring, S. Koh, T. Takayoshi, and N. Tanaka, Journal of Electron Microscopy 61, 17–23 (2012).